# An Anomaly-based Botnet Detection Approach for Identifying Stealthy Botnets

*Sajjad Arshad[1], Maghsoud Abbaspour[1], Mehdi Kharrazi[2], Hooman Sanatkar[1]*

[1]Electrical and Computer Engineering Department, Shahid Beheshti University; G.C; Tehran, Iran

[2] Department of Computer Engineering, Sharif University of Technology; Tehran, Iran

s.arshad@mail.sbu.ac.ir, maghsoud@ipm.ir, kharrazi@sharif.edu, hooman.sanatkar@mail.sbu.ac.ir

*Abstract*—Botnets (networks of compromised computers) are often used for malicious activities such as spam, click fraud, identity theft, phishing, and distributed denial of service (DDoS) attacks. Most of previous researches have introduced fully or partially signature-based botnet detection approaches. In this paper, we propose a fully anomaly-based approach that requires no a priori knowledge of bot signatures, botnet C&C protocols, and C&C server addresses. We start from inherent characteristics of botnets. Bots connect to the C&C channel and execute the received commands. Bots belonging to the same botnet receive the same commands that causes them having similar netflows characteristics and performing same attacks. Our method clusters bots with similar netflows and attacks in different time windows and perform correlation to identify bot infected hosts. We have developed a prototype system and evaluated it with real-world traces including normal traffic and several real-world botnet traces. The results show that our approach has high detection accuracy and low false positive.

## I. INTRODUCTION

Botnet is a collection of compromised hosts (bots) that are under control of an attacker (botmaster). Bots are used to send spam emails, host phishing web sites, cooperate in distributed denial of service (DDoS) attacks, and other kinds of malicious activities. Botmaster needs a command and control (C&C) channel to command the bots and coordinate malicious activities. Most of botnets C&C channels are using IRC (Internet Relay Chat) protocol. In this protocol, botmaster has a real-time communication with the bots. There are also a few botnets that use the HTTP protocol for C&C channels [1][2]. In HTTP-based C&C, the botmaster does not communicate directly with the bots. Instead, the bots periodically contact the C&C server to obtain their commands. These two protocols provide a centralized C&C mechanism. The main disadvantage of centralized C&C mechanism is the single-point-of-failure problem. For example, if IRC or HTTP server is taken down, the botmaster will not be able to communicate with the bots anymore. Thus, botmasters began using peer-to-peer (P2P) protocols for C&C channels. Currently, Storm Worm [3] and Nugache [4] are the most popular P2P botnets.

For designing a botnet detection approach that is resistant to the changes of the C&C mechanisms, we should study inherent characteristics of botnet behaviors. Bots connect to the C&C channel and execute the received commands. Bots belonging to the same botnet receive the same commands that causes them having similar netflows characteristics and performing same attacks. There are netflows that present communication between bots and C&C servers such as binary downloading and sending spam. Also, there are several types of attacks like scanning and distributed denial of service (DDoS). Our method is based on detecting these anomalous behaviors and finding meaningful relation between these activities to detect set of bots inside the monitored network. This paper makes the following main contributions:

- We propose an anomaly-based method that requires no a priori knowledge of bot signatures, botnet C&C protocols, and the C&C server addresses.
- We illustrate a method that can detect bots in the monitored network in real-time.
- In addition to detect bots with malicious activities (e.g. scanning, DDoS), our method can detect bots that do not perform malicious activities. But, our approach detects bots with malicious activities so fast.
- We have developed a prototype system based on our method and evaluated it with real-world network traces including normal traffic and several real-world botnet traces.

The rest of the paper is organized as follows. We review the related works in section 2. Section 3 describes the architecture and implementation of the approach. Section 4 evaluates the effectiveness of the approach on various network traces and finally the paper is concluded in section 5.

## II. RELATED WORKS

There are several researches that propose different botnet detection approaches. Binkley and Singh [5] combines IRC statistics and TCP work weight to detect IRC-based botnets. Karasaridis [6] used IRC netflows and scanning activities to detect IRC botnet controllers. Livadas [7] proposed a machine learning based approach which uses network-level traffic features of chat protocols for botnet detection. Rishi [8] is a signature-based IRC botnet detection approach that finds suspicious nicknames by analyzing IRC-based botnets traffic. The above approaches are used for only detecting IRC-based botnets; whereas we do not have assumption about botnet C&C protocols. BotHunter [9] uses a user-defined bot infection model in order to correlate alerts (e.g. scanning). BotSniffer [10] is a centralized botnet C&C detection approach that performs group analysis across multiple hosts through different time windows. But, it is used for detecting centralized botnets. BotMiner [11] clusters hosts based on their netflow-level statistics (e.g., the byte count, the packet count, etc.) and their malicious activities (e.g., scanning) and then correlates clusters to find bots. Nevertheless, it works offline and cannot detect bots quickly. BotProbe [12] proposes an active botnet probing technique that detects bots by distinguishing botnet C&C dialogs from human-human conversations. The aforementioned approaches are mostly signature-based and cannot find unknown botnets.

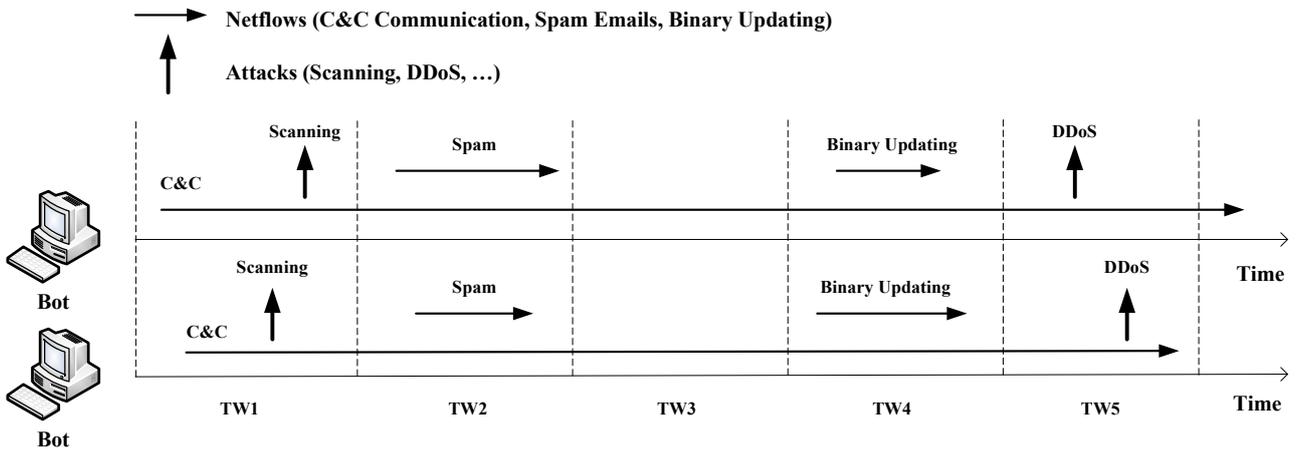

Figure 1. Similarity in bots behaviors in consecutive time windows

III. BOTNET DETECTION APPROACH

In this section, we illustrate a new botnet detection approach whose goal is to detect groups of bot infected hosts which belong to the same botnet in a monitored network. Bots belonging to the same botnet have similar behaviors during a time window and this similarity continues during consecutive time windows as shown in Fig. 1. Our approach finds behavior similarity of hosts in different properties such as netflow information through a predefined time window and tries to detect bots by correlating these similar behaviors between different time windows.

Fig. 2 shows the architecture of our botnet detection approach, which consists of nine interconnected components that analyze traffic online. Traffic dispatching component delivers traffic to Domain-IP Mapping, Netflow Generating and Alert Generating components. Domain-IP Mapping component maps the DNS domains to corresponded IPs for filtering purposes. Netflow Generating component generates TCP netflows between hosts. Alert Generating component reports the malicious activities of the hosts like scanning. The following five components perform at the end of each time window. Alert Filtering component filters useless alerts generated by Alert Generating component and Netflow Filtering filters the netflows generated by Netflow Generating component by using database produced by Domain-IP Mapping component. Then Netflow Clustering and Alert Clustering components cluster non-filtered netflows and alerts. Finally at the end of each time window correlation engine correlates the generated alert clusters and netflow clusters in order to detect bot infected hosts. The following sections illustrate architecture of our approach in detail. Our botnet detection approach operates on online traffic and detects bots in real-time. The following sections will illustrate each component in details.

*A. Traffic Dispatching*

This component has been designed for separating different types of traffic of Domain-IP Mapping, Netflow Generating and Alert Generating components. It delivers DNS traffic to Domain-IP Mapping and TCP traffic to Netflow Generating component. Also TCP, UDP and ICMP traffic is delivered to Alert Generating component. Irrelevant traffics like DHCP and ARP will be filtered out in order to increase performance.

*B. Domain-IP Mapping*

As discussed above, the bots can use the IPs or DNS names hard coded in their binary file in order to connect to the C&C servers. This component has been designed to map the domains and IPs to each other. This component processes DNS queries and responses in order to store in database the IP addresses in correspondence with the queried domain names. This database only is used to filter netflow to the sites in the white list as described in Netflow Filtering section.

*C. Netflow Generating*

This component processes TCP traffic and generates the netflows between hosts. Most of the routers like Cisco and Juniper have the capability of generating netflows. Also open source tools like ARGUS exist. We developed our own efficient tool to generate netflows. In addition to simple netflows, our tool is able to generate partial netflows (netflows that is not finished inside a time windows, is considered as a completed netflow, the remained part of the netflow is considered as another netflow in next time windows). The most important property of our tool is its efficiency and customized netflow record generating. Currently, we consider only TCP netflows. Each netflow has the following information: start time, end time, source

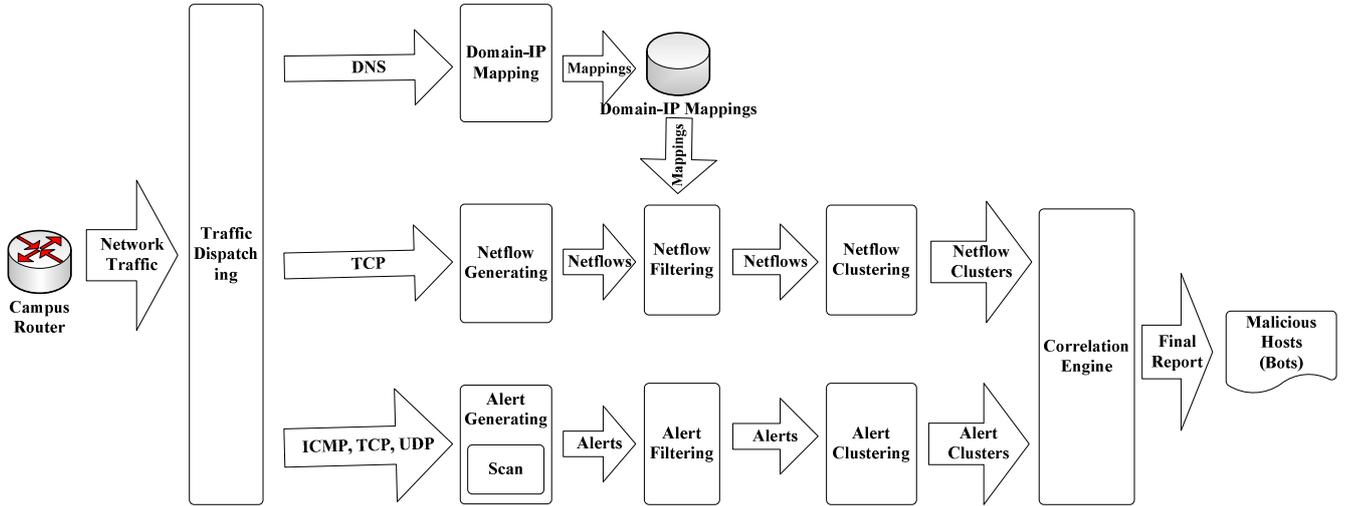

Figure 2. Botnet Detection Approach Architecture

IP/Port, destination IP/Port, number of sent/received packets and number of sent/received bytes.

### D. Alert Generating

This component reports the malicious activities of the hosts like scanning which is the most important activity that can be done by bots. Currently, this component contains sub-component scanning which will be illustrated in the following. The scanning sub-component uses the sfPortscan preprocessor of Snort (an open source IDS) which can detect several types of scanning such as decoy portscan, distributed portscan and portsweep for TCP, UDP and ICMP protocols.

### E. Netflow Filtering

In order to increase the performance of the approach, we filter out useless netflows by the following rules:

1. Netflows related to communications between internal hosts and netflows initiated from external hosts towards internal hosts. Netflows of which do not contain any payload (e.g. scanning netflows).
2. Netflows of which contain bulky payloads that seem to be large file downloading traffic and their payloads size are higher than BULKY_THR factor. In our experiment, we consider it as 1MB. By analyzing the traffic of several botnet types we found out that the maximum netflow size of them is lower than this threshold.
3. Netflows of which their destination server is in the whitelist (known servers like google and yahoo). This list is generated and updated by system administrator manually.

This filtering does not have any effect on the detection accuracy of our approach and only improves the efficiency of our approach.

### F. Alert Filtering

This component filters out useless alerts generated by activities such as scanning by external hosts. This filtering does not have any effect on the detection accuracy of our approach and only improves the efficiency of our approach.

### G. Netflow Clustering

This component clusters the netflows with similar communication patterns in the given time window based on flow-level feature set and payloads. Fig. 3 shows this two-level netflow clustering.

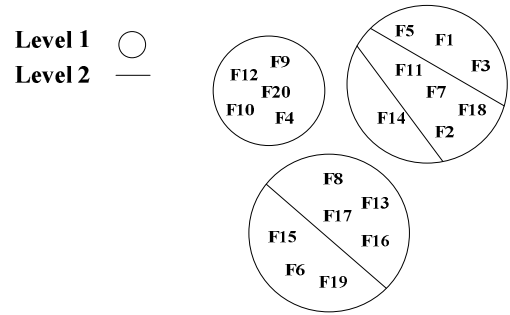

Figure 3. Two-Level Netflow Clustering

In the first level clustering, we determine a number of features for each netflow that together create a vector for that netflow. These features are: number of sent packets, number of sent bytes, average number of bytes per sent packet, average number of sent bytes per second, number of received packets, number of received bytes, average number of bytes per received packet, and average number of received bytes per second.

The netflow vectors of two bots in a botnet are similar to each other; therefore we use a clustering algorithm in order to cluster similar vectors. For first level clustering, X-means clustering [13] method has been used which in contrast to K-means algorithm does not need to know the number of clusters. X-means repeats K-means cycle and uses Bayesian Information Criterion [13] to calculate the best value for K.

In order to increase the accuracy of clustering, we perform second level clustering on the first level generated clusters. As a result, first level clusters are broken to new clusters in which the single netflow clusters are filtered out. Second level clustering performs on the payloads of the netflows. Each netflow has two payloads: the sent payload which is obtained by joining sent packets from the source to the destination, and the received payload which is obtained by joining the sent packets from the destination to the source. For second level clustering, we use hierarchical clustering, and NCD [14] is used for calculating the payload similarity. Equation (1) is used for calculating the distance of a pair of netflows, which $w_s$ and $w_r$ coefficients are calculated using (2) and (3). In (1), $w_s + w_r = 1$ and $SP$ is the sent payload and $RP$ is the received payload.

$$d(F_i, F_j) = w_s \times NCD(SP_i, SP_j) + w_r \times NCD(RP_i, RP_j) \quad (1)$$

$$w_s = \frac{|SP_i| + |SP_j|}{|SP_i| + |SP_j| + |RP_i| + |RP_j|} \quad (2)$$

$$w_r = \frac{|RP_i| + |RP_j|}{|SP_i| + |SP_j| + |RP_i| + |RP_j|} \quad (3)$$

We calculate the distance between each pair of netflows and then form a hierarchical tree. Therefore, we have to extract the clusters out of this tree. Number of these clusters depends on cut-off factor $DIST\_THR$ which shows the distance between netflow payloads. By analyzing the netflows of different types of botnets and calculating their payload distance we consider this factor as 0.35.

### H. Alert Clustering

This component performs a clustering on the generated scanning alerts. We can use the following features for clustering of scanning alerts: scanned port, scanned net or subnet, and type of scanning. In this implementation, type of scanning has been used for clustering of scanning alerts.

### I. Correlation Engine

This component considers several time windows that their size ($TW\_SIZE$) which its default value is 20 minutes. We found that different types of botnets have at least one activity during his time. Correlation operation is performed at the end of each time window. Fig. 4 shows the netflow and the alert clusters generated in the time windows.

The correlation engine component has a table of hosts in which every host has a specific score. Now, during the correlation step, this component correlates the clusters inside the current time window with all other clusters in the same time window, and with the clusters of a number ($MAX\_NUM\_TW$) of previous time windows. Therefore, if bots are active in a time window, they will be detected at the end of that time window. Its default value is set to 3 because maximum distance between bots activity is at most 3 time windows.

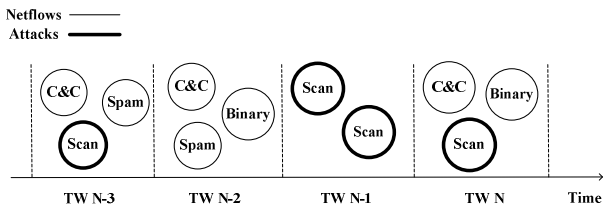

Figure 4. Generated clusters in the time windows

The score of the hosts which belong to the correlated clusters is increased, and the score of the hosts which did not belong to the correlated clusters is decreased. Fig. 5 shows the algorithm which calculates the score of host $h$ in time window $N$. This score is calculated as the maximum return value of the function $TWCorrelation$ (presented in Fig. 6) in $MAX\_NUM\_TW$ iterative calls (for $MAX\_NUM\_TW$ previous time windows). Also, the score of the hosts which did not belong to the correlated clusters is decreased by difference of current time window and the last correlated time window ($lastCorrelatedTW$) of those hosts.

```
function Score(h, N):
    maxScore = −1

    for i = 0 to MAX_NUM_TW:
        twScore = TWCorrelation(h, TW_N, TW_{N−i})
        if twScore ≥ 0:
            if maxScore == −1 or twScore > maxScore:
                maxScore = twScore
            endif
        endif
    endfor

    if maxScore ≠ −1:
        return maxScore
    else:
        return − 1 × (N − lastCorrelatedTW_h)
end function
```

Figure 5. Score of host h in Time Window N

Fig. 6 shows the function that calculates correlation score between two time windows ($TWCorrelation$).

```
function TWCorrelation(h, TW_i, TW_j):
    clusterSet = {}
    twScore = 0
    for c_i in C^h_{TW_i}:
        for c_j in C^h_{TW_j}:
            if ClusterCorrelation(c_i, c_j) ≥ CORR_THR:
                clusterSet.insert(c_i)
                clusterSet.insert(c_j)
            endif
        endfor
    endfor

    for c in clusterSet:
        twScore += {1,  type(c) == Netflow
                    {2,  type(c) == Scanning
    endfor

    return min(twScore, MAX_TW_SCORE)
end function
```

Figure 6. Time window correlation

Equation (4) shows the correlation degree calculation method. In (4), $c_i$ and $c_j$ are two clusters and $j \geq i$.

$$ClusterCorrelation(c_i, c_j) = 1 - e^{-\left(\frac{|c_i \cap c_j|}{|c_i \cup c_j|} \times |c_i \cap c_j| \times \frac{1}{(j-i)+1}\right)} \quad (4)$$

Correlation degree is close to 1 if the two clusters have higher intersection percent, higher intersection count, and less time window distance (distance between time windows which the two clusters belong to).

The TWCorrelation function considers all of the correlated clusters. Each netflow cluster adds 1, and each alert cluster adds 2 points to the time window correlation score. To reduce false positives, the maximum calculated score of a host in each time window is MAX_TW_SCORE. We set this factor to 5. We call two clusters correlated, if their correlation degree exceeds CORR_THR (correlation threshold). A host is reported as a bot, if its score exceeds BOT_THR (bot threshold). By studying different types of botnets, in our experiment, CORR_THR and BOT_THR is considered 0.65 and 33 respectively. In the following

TABLE I.
NETWORK TRACES STATISTICS

| Trace | Num. of Clients | Size | Duration | Num. of Packets | | | | Num. of TCP Netflows |
|---|---|---|---|---|---|---|---|---|
| | | | | Total | ICMP | TCP | UDP | |
| Normal | 28 | 44 GB | 215 h | 74,974,692 | 658,747 | 61,752,609 | 12,563,336 | 818,362 |
| IRC-SdBot | 4 | 2.3 MB | 30 h | 18,685 | 0 | 17,072 | 1,613 | 16 |
| IRC-SpyBot | 4 | 37.9 MB | 30 h | 468,026 | 2,708 | 464,054 | 1,264 | 421,870 |
| HTTP-Bot-I | 4 | 6.2 MB | 30 h | 16,926 | 0 | 15,214 | 1,712 | 1,164 |
| HTTP-Bot-II | 4 | 5.3 MB | 30 h | 14,594 | 0 | 14,054 | 540 | 1,069 |

section, we evaluate the effectiveness of our approach on various network traces.

## IV. EXPERIMENTS

To evaluate the performance of proposed botnet detection approach, it is tested on several real-world network traces, including normal and collected botnet traffics.

### A. Data Collection

We have captured part of our traffic traces for continuous 8 days period from part of our campus network. We assume that this traffic trace is clean because of the two following reasons. First, the traffic trace has been captured from the network which is behind NAT and the outside attackers cannot exploit internal hosts. Second, internal network users do not have administrator access and cannot run malicious binaries (e.g. downloading and running suspicious email attachments). A random sampling of the network trace shows that the traffic is diverse, containing normal application protocols, such as HTTP, FTP, SSH, DNS. We have collected a total of four different botnets covering IRC and HTTP. The basic statistics of these traces are listed in Table I. We use two IRC and two HTTP botnet traces, i.e., IRC-SdBot, IRC-SpyBot, HTTP-Bot-I and HTTP-Bot-II. IRC-SdBot and IRC-SpyBot are generated by executing modified bot code (SdBot and SpyBot [15]) in a fully controlled virtual network. Each trace contains four Windows XP IRC bot clients, and last 30 hours. HTTP-Bot-I and HTTP-Bot-II are generated according to the description of web-based C&C communications in [1][2]. In HTTP-Bot-I, the bot periodically connect to HTTP server every five minutes and the whole trace lasts for 30 hours. In HTTP-Bot-II, we have a more stealthy C&C communication where the bot waits a random time between 0 to 10 minutes each time before it visits the server and the whole trace lasts for 30 hours. The bots communicate with a controlled C&C server (IRC or HTTP server) and execute the received commands.

We set up a virtual network environment with 17 Windows XP virtual machines using VirtualBox. We ran each bot on four Windows XP virtual machines and an IRC server and an HTTP server on one Windows XP virtual machine.

### B. Evaluation Results

In order to validate the detection accuracy of our approach, we mixed botnet traffic to normal traffic. We also initially did not apply the forth filtering rule listed in section Netflow Filtering (whitelist) on traffic traces.

Table II reports the detection results for each botnet. Table II shows that our approach is able to detect all four botnets. For 3 out of 4 botnets, we obtained 100% detection rate, i.e., we successfully identified all the bots within the 3 botnets. For example, in the case of IRC botnets (SdBot and SpyBot), our approach correctly detected all of the botnet members. HTTP-Bot-I contains 4 bot clients.

TABLE II.
DETECTION RESULTS

| Trace | Num. of Clients | Detected as Bot | FP/FN |
|---|---|---|---|
| Normal | 28 | 2 | 2/0 |
| IRC-SdBot | 4 | 4 | 0/0 |
| IRC-SpyBot | 4 | 4 | 0/0 |
| HTTP-Bot-I | 4 | 4 | 0/0 |
| HTTP-Bot-II | 4 | 2 | 0/2 |

Our approach was able to identify all of the bots. In the case of HTTP-Bot-II, we correctly detected bots. However, two of bots belonging to the botnet was not detected, which means that the detector generated two false negatives. There were some cases in which our approach also generated two false positives. We discovered that by applying the forth filtering rule of section 3.5 which use white list, the false positives no longer occur.

As we can see, our approach performs very well in our experiments, showing very high detection accuracy with relatively low false positives in real-world network traces.

## V. CONCLUSION & FUTURE WORK

Botnet detection is a very challenging research area. In this paper, we presented a fully anomaly-based botnet detection approach that requires no a priori knowledge of bot signatures, botnet C&C protocols, and C&C server addresses. Our approach is based on the intuition that bots respond to commands and perform malicious activities in a similar fashion. Our approach shows high detection accuracy on real-world botnets with a low false positive on normal traffic.

In our future work, we will study DDoS attacks and implement a DDoS alert generating sub-component for our alert generating component. In addition, we plan to further improve the efficiency of the netflow clustering algorithm to work in very high speed networks.

ACKNOWLEDGMENTS

This work is supported by Iranian Education and Research Institute for Information and Communication Technology (ERICT) under grant 500/8974.